\documentclass[prd, aps, twocolumn, superscriptaddress, amsmath, amssymb, showkeys]{revtex4-2}
\usepackage[utf8]{inputenc}
\usepackage{textgreek}
\usepackage{natbib}
\usepackage{booktabs}
\usepackage{graphicx}
\usepackage{dcolumn}
\usepackage{bm}
\usepackage{amsmath}
\usepackage{multirow}

\graphicspath{{figures/}}

\usepackage[colorlinks=true,linkcolor=red,urlcolor=blue,citecolor=blue]{hyperref}

\begin{document}

\title{Relativistic Corrections to the Formation Rate of Extreme Mass-Ratio Inspirals}

\author{Chen Feng}
\affiliation{School of Astronomy and Space Sciences, University of Chinese Academy of Sciences (UCAS), Beijing 100049, China}

\author{Yong Tang}
\affiliation{School of Astronomy and Space Sciences, University of Chinese Academy of Sciences (UCAS), Beijing 100049, China}
\affiliation{School of Fundamental Physics and Mathematical Sciences, Hangzhou Institute for Advanced Study, UCAS, Hangzhou 310024, China}
\affiliation{International Center for Theoretical Physics Asia-Pacific, Beijing/Hangzhou, China}

\begin{abstract}
Extreme mass-ratio inspirals (EMRIs) are long-duration gravitational-wave sources in which a compact object gradually spirals into a massive black hole. Their formation is governed by the interplay between stochastic angular-momentum diffusion driven by two-body relaxation and the dissipative evolution caused by gravitational-wave emission, with the loss-cone boundary deciding whether an object undergoes an inspiral or a direct plunge. Building on this physical picture, we construct a relativistically self-consistent analytic framework for estimating EMRI event rates. In Schwarzschild spacetime, we generalize the standard loss-cone angular momentum to an energy-dependent quantity and revise the plunge pericenter by using the minimum stable radius derived from general relativity. Relative to the Newtonian treatment, we show that incorporating these relativistic effects increases the predicted EMRI rates by roughly a factor of 8. This enhancement becomes more pronounced for shallower stellar density profiles and is insensitive to the mass of the central massive black hole, which emphasizes that relativistic effects are essential for EMRI rate estimations that are relevant for space-based gravitational-wave detectors, such as LISA and Taiji.
\end{abstract}


\maketitle

\section{Introduction}\label{sec:introduction}
Extreme mass ratio inspirals (EMRIs) are long-lived gravitational-wave (GW) sources in which a compact object (CO) of mass $m$—typically a stellar-mass black hole, neutron star, or white dwarf—inspirals into a massive black hole (MBH) of mass $M_{\mathrm{BH}}\gg m$~\cite{Finn:2000sy,eLISA:2013xep,Oliver:2023xan,Rom:2024nso,Pan:2021lyw,Zhao:2025hqd}. Because the small body completes orbital cycles $\sim 10^{4}$--$10^{6}$ deep in the potential of the MBH before merger, EMRIs encode detailed information about spacetime geometry~\cite{Zhang:2024ogc,Fu:2024cfk,Cardenas-Avendano:2024mqp,Shen:2025svs,Liu:2025sgz}, the mass and spin of the MBH~\cite{Barack:2006pq,Babak:2006uv,Amaro-Seoane:2012jcd,Liang:2022gdk,Yang:2024cnd,Zou:2025fsg}, and the dense environment around galactic nuclei~\cite{Dai:2023cft,Li:2021pxf,Zhang:2024hrq,Alexander:2017rvg,Yue:2018vtk,Zhang:2022rfr,Zhang:2024ugv,Peng:2024wqf,Peng:2022hqa,Yue:2018vtk,Wang:2025rrj,Wang:2025msq}. Their characteristic frequencies lie in the millihertz band, making them great targets for space-based GW observatories such as LISA~\cite{Barausse:2020rsu,Danzmann:1994mma,Barack:2003fp,Amaro-Seoane:2007osp,Babak:2017tow} and Taiji~\cite{Ruan:2018tsw,Hu:2017mde}.

There are several channels to form EMRIs~\cite{Chen:2018axp, Pan:2021oob, Mancieri:2024sfy, Rom:2025ppy, Yin:2024nyz, Cui:2025bgu}. Here we focus on the scenario that the formation of EMRIs is governed by the interplay between orbital diffusion due to two-body relaxation and dissipative evolution driven by GW emission~\cite{Amaro-Seoane:2012lgq}. Far from the MBH, orbits are approximately Keplerian, and their angular momenta undergo stochastic diffusion in phase space. As the angular momenta undergo random walks toward smaller values, the periapsis can move sufficiently close to the MBH that GW emission becomes the primary cause of orbital evolution, progressively circularizing and tightening the orbit until the compact object inspirals and merges. However, if the angular momentum drops below the critical “loss-cone” threshold before GW emission takes over, the object plunges directly into the MBH. The event rate of EMRIs, therefore, depends sensitively on the competition between relaxation and GW emission, and on the precise definition of the loss-cone~\cite{Amaro-Seoane:2012lgq}.

The event rate of EMRI is a key physical quantity that affects the overall strength of gravitational-wave signals of EMRI in LISA and Taiji~\cite{Bonetti:2020jku,Seoane:2025gfh,Seoane:2024nus}. This study adopts an analytic framework based on the EMRIs formation picture combined with a one-dimensional Fokker–Planck treatment~\cite{Mancieri:2025cmx,Broggi:2022udp,Lightman:1977zz, Hopman:2006xn,Hopman:2006qr,Amaro-Seoane:2010dzj,Bar_Or_2016,Pan:2021ksp,Broggi_2022,Broggi:2024mww,Zhang:2025jmm,Merritt:2015vxa,Merritt:2015xpa,Merritt:2015elb,Merritt:2015kba}: assuming an approximately isotropic stellar distribution, the steady-state flux into the loss-cone is governed by angular-momentum diffusion, and its magnitude is jointly determined by the number density on the energy shell, the local relaxation time, and a logarithmic factor characterizing the efficiency of crossing the loss-cone boundary. By mapping energy to the semi-major axis, one can integrate this flux over the semi-major axis interval that permits EMRIs formation to obtain the total event rate.

We construct a relativistically consistent analytical model for EMRI event rates grounded in the physical picture of EMRI formation, and provide a quantitative comparison with the Newtonian case. We generalize the loss-cone angular momentum from a constant to an energy-dependent function in Schwarzschild spacetime and, via the relaxation–gravitational-wave matching curve, map it to a function of the initial semi-major axis. We revise the plunge pericenter using the minimum stable radius in general relativity (GR), thereby redefining the critical semi-major axis and the loss-cone boundary. Under these relativistic modifications, we compute the total EMRIs event rate and compare it with the conventional Newtonian approximation, quantifying the impact on the rate and its dependence on the stellar density slope and the MBH mass. We find that employing the relativistic loss-cone angular momentum and the plunge pericenter increases the EMRIs rate by a factor of about $8$. Moreover, this factor increases with decreasing density slope and is insensitive to the MBH mass. This implies that relativistic effects are essential for reliable EMRI rate forecasts in the scientific objectives of future space-based gravitational-wave detectors, such as LISA and Taiji.

The paper is organized as follows: Sec.~\ref{sec:Discussion of event rate} reviews the EMRIs event-rate framework based on angular-momentum diffusion, establishes the steady-state stellar flux into the MBH, and defines the critical semi-major axis and the loss-cone boundary; Sec.~\ref{sec:GR_Loss_Cone_Boundary} refines the GR loss-cone boundary in Schwarzschild and evaluates its impact on $a_{\mathrm{cri}}$ and the loss-cone boundary; Sec.~\ref{sec:Impact_on_EMRIs_event_rate} calculates the EMRIs rate, compares the GR results with those under the Newtonian approximation, and analyzes the dependence on the cusp slope and the MBH mass; Sec.~\ref{sec:conclusion} presents the conclusions.

\section{The event rate of EMRIs}\label{sec:Discussion of event rate}
In this section, we provide a concise review of EMRI formation processes, the theoretical framework, and the associated analytical formulations for the EMRI event rate, and examine their implications. Subsequently, we identify the physical parameters that may affect the event rate.

\subsection{The formation of EMRIs}

An MBH typically hosts a dense stellar cusp that acts as the cradle for EMRIs formation, with a profile depending on distance from the MBH center $r$, $\rho \propto r^{-\gamma}$. In most cases, the slope of the cusp $\gamma$ is shallower than 2. Observations of nearby galactic nuclei find $\gamma \sim 1-1.5$~\cite{G_ltekin_2009}, while numerical simulations suggest that two-body relaxation in isolation produces $\gamma \approx 7/4$ (the Bahcall–Wolf value), but various astrophysical processes—including binary stellar evolution, massive black hole binaries, and galaxy mergers—can flatten the profile to $\gamma < 2$~\cite{Alexander:2017rvg,Merritt:2013awa}. Within a spherical shell that spans from $r$ to $r + dr$, the number of compact objects (COs) is proportional to $4 \pi r^2\rho$. Consequently, the number of COs generally increases with $r$. At very small $r$, COs are rapidly depleted and cannot maintain a steady supply, so EMRIs are predominantly produced at large radii.
  
At large radii, GW emission is too weak for the orbit to decay appreciably. COs in the stellar cusp undergo relaxation driven by interactions with other COs. Relaxation can drive the orbital eccentricity $e$ to large values while keeping the semi-major axis $a$ almost constant, thus reducing the periapsis to a distance from the MBH where GW emission becomes significant. Once GW emission dominates, the orbit decays and the CO eventually merges with the MBH. This pathway forms EMRIs. 

On the onset of GW's dominance, the orbit of the test object can still be approximated as an elliptical Keplerian orbit. Consequently, GW emission drives the orbital evolution with time $t$~\cite{Peters:1963ux}
\begin{align}
&\left\langle \frac{da}{dt} \right\rangle_{\rm{gw}} = 
-\frac{64}{5} \frac{G^3 \mu M^2}{ a^3 (1 - e^2)^{7/2}} 
\left( 1 + \frac{73}{24} e^2 + \frac{37}{96} e^4 \right),
\end{align}
\begin{align}
&\left\langle \frac{de}{dt} \right\rangle_{\rm{gw}} = 
-\frac{304}{15} \frac{e \, G^3 \mu M^2}{ a^4 (1 - e^2)^{5/2}} 
\left( 1 + \frac{121}{304} e^2 \right).
\end{align}
Here $G$ is the Newton's constant, and we have set $c=1$. We consider a binary system with component masses $m$ and $M_{\mathrm{BH}}$, which, therefore, has total mass $M = m + M_{\mathrm{BH}}$ and reduced mass $\mu = m M_{\mathrm{BH}} / M$.

Not all test objects can form EMRIs. Moreover, the larger the semi-major axis of the test object, the smaller the required periapsis for GW emission to become significant. However, if the semi-major axis is sufficiently large such that the requisite periapsis is so small as to effectively correspond to a merger with the MBH, the system cannot form EMRIs and instead undergoes a direct plunge. We use the critical semi-major axis $a_{\mathrm{cri}}$ to denote the maximum distance at which EMRIs can form.

For the above reasons, there is an upper limit to the semi-major axis of the formation of EMRIs~\cite{Amaro-Seoane:2012lgq} deterimined by
\begin{equation}\label{eq:condition_to_form_emir}
    \begin{aligned}
        t_{\mathrm{gw}} &= t_{\mathrm{rlx};\, r_{p}}, \\
        a(1 - e) &= r_{\mathrm{p};\, \mathrm{lc}}.
\end{aligned}
\end{equation}
Here, $t_{\mathrm{gw}}$ and $t_{\mathrm{rlx};\, r_{p}}$ denote the characteristic timescales for the evolution of periapsis due to the emission of GW and relaxation, respectively. The quantity $r_{\mathrm{p};\,\mathrm{lc}}$ denotes the minimum periapsis distance at which an orbit can exist. The subscript `$\mathrm{lc}$' denotes the critical value for capture by the MBH, \textit{i.e.}, the boundary of the loss-cone region, which is defined by orbits sufficiently close to the massive black hole to be removed by capture. $t_{\mathrm{gw}}$ and $t_{\mathrm{rlx};\, r_{p}}$ are given by~\cite{Amaro-Seoane:2019umn}
\begin{align}
    t_{\mathrm{gw}} = \frac{1 - e}{[d(1 - e)/dt]_{\rm{gw}}} = -\frac{1 - e}{(de/dt)_{\rm{gw}}},
\end{align}
\begin{align}
    t_{\mathrm{rlx};\, r_{p}} = \frac{0.34 \sigma_{\rm{f}}^3}{G^2 m_{\rm f} \rho_{\rm f} \ln \Lambda'}(1 - e),
\end{align}
where $\sigma_{\rm{f}}$ is the velocity dispersion of field objects, $m_{\rm f}$ is their mass, $\rho_{\rm f}$ is the density, and $\ln \Lambda' = \ln(M_{\rm BH}/m_f)$ is the Coulomb logarithm. 

We model the stellar density around a massive black hole using a power-law profile normalized by a characteristic radius defined from the velocity dispersion. The velocity dispersion outside the influence radius is taken to scale with black-hole mass as~\cite{G_ltekin_2009}
\begin{align}
    \sigma_{\mathrm{0}} \;=\; 70~\mathrm{km\,s^{-1}}\,
    \left(\frac{M_{\mathrm{BH}}}{1.53\times 10^{6}\,M_{\odot}}\right)^{1/4.24},  
\end{align}
where $M_{\odot}$ is the solar mass. In the MBH influence radius, $\sigma_{\mathrm{f}}^2 = {G\,M_{\mathrm{BH}}}/{r}$~\cite{Amaro-Seoane:2012lgq}. The characteristic (influence) radius is then defined as
\begin{align}
    r_{\mathrm{h}} \;=\; \frac{G\,M_{\mathrm{BH}}}{\sigma_{\mathrm{0}}^{2}}.
\end{align}

Within $r \lesssim r_{\mathrm{h}}$, we consider a single power-law stellar density profile
\begin{align}
    \rho_{\mathrm{f}}(r) \;=\; \rho_{0}\,\left(\frac{r}{r_{\mathrm{h}}}\right)^{-\gamma},
\end{align}
with a constant slope index $\gamma$. Here, $\rho_{0}$ is a normalization factor determined by
\begin{align}
    \rho_{0} \;=\; \frac{(3-\gamma)\,M_{\mathrm{BH}}}{4\pi\,r_{\mathrm{h}}^{3}}.
\end{align}
Then the density profile can be written as
\begin{align}
    \rho_{\mathrm{f}}(r) \;=\; \frac{(3-\gamma)\,M_{\mathrm{BH}}}{4\pi\,r_{\mathrm{h}}^{3}}\,
    \left(\frac{r}{r_{\mathrm{h}}}\right)^{-\gamma}.
\end{align}

\begin{figure}[t]
    \centering
    \includegraphics[width=1\linewidth]{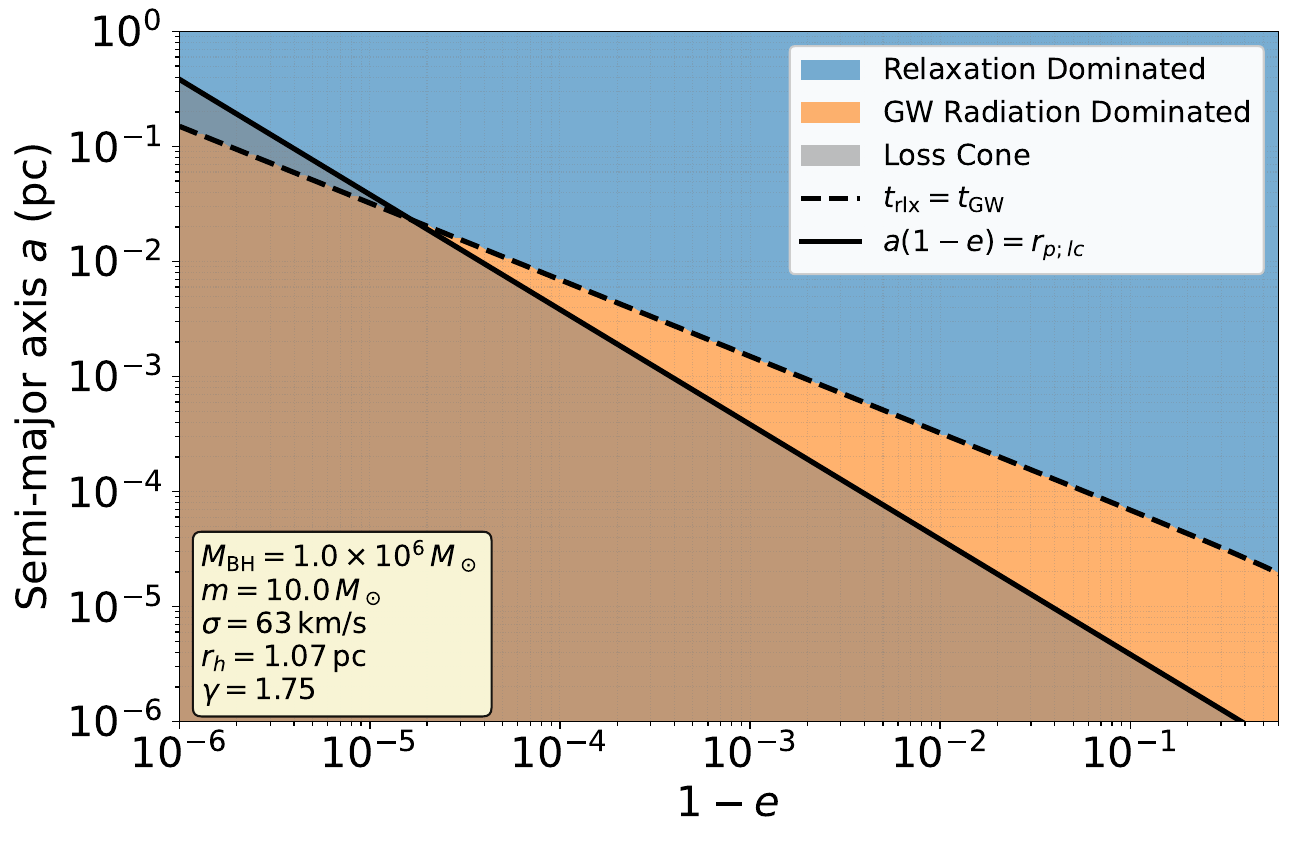}
    \caption{Effect-dominated region and loss-cone curve in the $(1-e)$--$a$ plane. The blue, orange, and gray regions correspond to the relaxation-dominated, GW-emission-dominated, and loss-cone regions, respectively. The solid line indicates $t_{\mathrm{gw}} = t_{\mathrm{rlx}}$, whereas the dashed line represents the loss-cone condition $a(1 - e) = r_{\mathrm{p};\, \mathrm{lc}}$.}
    \label{fig:emriform}
\end{figure}

As shown in Fig.~\ref{fig:emriform}, the dashed line specified by $t_{\mathrm{gw}} = t_{\mathrm{rlx}}$ intersects the boundary of the loss-cone region, thus defining $a_{\mathrm{cri}}$. If the initial semi-major axis exceeds $a_{\mathrm{cri}}$, the test object directly enters the loss-cone region and plunges into the MBH. Conversely, if it is smaller than $a_{\mathrm{cri}}$, the object enters the GW emission dominated region and forms an EMRI with the MBH.

\subsection{EMRIs event rate}
Now we consider the event rate of EMRIs through relaxation-driven diffusion in phase space. During relaxation, the stellar distribution function $f(E,L,t)$ executes a random walk in energy $E$ and angular momentum $L$ due to numerous weak gravitational encounters, which typically perturb $L$ more efficiently than $E$. The gradual diffusion to low $L$ is described by a one-dimensional orbit-averaged Fokker–Planck equation for $L$ at fixed $E$~\cite{Merritt:2013awa,Hopman:2005vr},

\begin{align}
     \frac{\partial N}{\partial t} = - \frac{1}{2}\frac{L_{\mathrm{m}}^2}{t_{\mathrm{rlx}}}\frac{\partial}{\partial L}\left\{  \frac{N}{L}  
    -  \frac{\partial  N}{\partial L}  \right\}.  
\end{align}
Here, $L_{\mathrm{m}}$ denotes the maximum angular momentum at $E$, i.e., the circular angular momentum $\sqrt{G M_{\mathrm{BH}} a}$.  $N(L,E,t)$ represents the stellar number density in phase space and is related to the phase-space distribution function $f(L,E,t)$ through $N(L,E,t)\,\mathrm{d}E\,\mathrm{d}L = 8\pi^{2} L\,P(E,L)\,f(L,E,t)\,\mathrm{d}E\,\mathrm{d}L$~\cite{Merritt2013-dr}, where $P(E, L)$ denotes the radial orbital period. $t_{\mathrm{rlx}}$ denotes the relaxation time~\cite{Amaro-Seoane:2012lgq}
\begin{equation}
    t_{\mathrm{rlx}} = \frac{t_{\mathrm{rlx};\, r_{p}}}{1-e} .
\end{equation}

The relaxation effect diffuses stellar orbits into the loss-cone region. Under steady-state conditions, the stellar current $\mathcal{F}$, which is solely a function of energy, denotes the flux of stars entering the loss cone per unit interval of energy. The current $\mathcal{F}$ is related to the change rate of the number of stars in the angular momentum space between $L_{\mathrm{lc}}$ and $L_{\mathrm{m}}$~\cite{Merritt:2013awa}:
\begin{align}\label{eq:Stellar_Current_Definition}
    \mathcal{F}(E)\, dE = - \frac{\partial}{\partial t} \left[ \int_{L_{\textrm{lc}}}^{L_m} N(E,L)\, dL \right] dE.
\end{align}

Physically, $\mathcal{F}(E)$ measures how rapidly stars on a given energy shell flow into the loss-cone and are captured by the MBH. Substituting the Fokker–Planck equation into Eq.~\ref{eq:Stellar_Current_Definition} yields
\begin{align}\label{eq:Stellar_Current_Solution}
    \mathcal{F}(E) = \frac{1}{2} \frac{L_{\mathrm{m}}^2}{t_{\mathrm{rlx}}} \left[ \frac{N(E, L)}{L} - \frac{\partial N(E, L)}{\partial L} \right] \bigg|_{L_{\textrm{lc}}}^{L_m}.
\end{align}
In $(E,L)$ phase space, the stellar number density can be written as~\cite{Hopman:2005vr}
\begin{align}\label{eq:N_E_L_Solution}
    N(E,L)=\frac{2\,N_{\mathrm{iso}}(E)\,L}{L_m^2(E)}\cdot
    \frac{\ln\!\left(L/L_{\textrm{lc}}\right)}{\ln\!\left(L_m/L_{\textrm{lc}}\right)}.
\end{align}
The factor $2N_{\mathrm{iso}}(E)L/L_m^2(E)$ represents the isotropic phase-space density in $(E,L)$; integrating it over $L$ from $0$ to $L_m$ gives the number density $N_{\mathrm{iso}}(E)$ at energy $E$. The logarithmic factor ensures that $N(E,L)$ satisfies the boundary conditions at $L_{\mathrm{lc}}$ and $L_m$: it vanishes at $L=L_{\mathrm{lc}}$ and matches the isotropic distribution at $L=L_m$. Substituting Eq.~\ref{eq:N_E_L_Solution} into Eq.~\ref{eq:Stellar_Current_Solution} yields the stellar current in the MBH per energy interval
\begin{align}
    \mathcal{F}(E) = \frac{N_{\mathrm{iso}}(E)}{\ln(L_{\mathrm{m}} / L_{\mathrm{lc}}) t_{\mathrm{rlx}}(E)}.
\end{align}

Considering Keplerian orbits, where energy and semi-major axes have a one-to-one correspondence, we can convert energy into the semi-major axis. This allows us to express the event rate as~\cite{Amaro-Seoane:2019umn}
\begin{align}\label{eq:evetrate}
    \Gamma = \int_{a_{\mathrm{min}}}^{a_{\mathrm{cri}}} \frac{da \, N_{\mathrm{iso}}(a)}{\ln(L_{\mathrm{m}} / L_{\mathrm{lc}}) t_{\mathrm{rlx}}(a)},
\end{align}
where
\begin{align}
    N_{\mathrm{iso}}(a)
    = \left[\,\frac{(3-\gamma)\,M_{\mathrm{BH}}}{r_h\,m_{\mathrm{f}}}\,\right]
    \left(\frac{a}{r_h}\right)^{2-\gamma},
\end{align}
$a_{\mathrm{min}}$ denotes the minimum distance at which EMRIs can form, and $a_{\mathrm{cri}}$ denotes the maximum distance. Since $a_{\mathrm{cri}}$ is determined by Eq.~\ref{eq:condition_to_form_emir}, which involves $r_{\mathrm{p};\,\mathrm{lc}}$, the value of $a_{\mathrm{cri}}$ depends on the loss-cone boundary defined by $L_{\mathrm{lc}}$ and $r_{\mathrm{p};\,\mathrm{lc}}$. These quantities will be examined in detail in the following section.

\section{General Relativistic Treatment of Loss-Cone Boundary}\label{sec:GR_Loss_Cone_Boundary}
In this section, we present a general relativistic treatment of the loss-cone boundary. We first derive the loss-cone angular momentum in Schwarzschild spacetime and show its dependence on orbital energy. We then apply this framework to the EMRI formation scenario, demonstrating how $L_{\mathrm{lc}}$ varies with the semi-major axis. Finally, we incorporate the resulting energy-dependent plunge pericenter into the loss-cone boundary and assess its impact on $a_{\mathrm{cri}}$.

\subsection{Loss-cone angular momentum in Schwarzschild spacetime}\label{subsec:Schwarzschild_Llc}
The loss-cone angular momentum $L_{\mathrm{lc}}$ represents the minimum angular momentum that the orbit can maintain. For a given energy $E$, $L_{\mathrm{lc}}$ depends on both $E$ and the MBH mass $M_{\mathrm{BH}}$. For a Schwarzschild MBH, the metric is
\begin{multline}
    d\tau^{2} = -\left(1 - \frac{2GM_{\mathrm{BH}}}{r}\right) dt^{2} + \left(1 - \frac{2GM_{\mathrm{BH}}}{r}\right)^{-1} dr^{2}  \\
    + r^{2} d\Omega^{2},
\end{multline}
where $d\Omega^{2} = d\theta^{2} + \sin^{2}\theta\, d\phi^{2}$. The equation of motion for a test object orbiting a Schwarzschild MBH is

\begin{align}\label{eq:EOM-schwarzshild}
    \left( \frac{dr}{d\tau}\right)^{2} &= E^2 - \left(1-\frac{2GM_{\mathrm{BH}}}{r}\right)\!\left(1+\frac{L^2}{r^2}\right) \notag \\
     &=E^2 - V(r).
\end{align}
Here, $V(r)$ denotes the effective potential. For a given energy $E$, different angular momenta yield different shapes of the effective potential. When the angular momentum equals the critical angular momentum $L_{\mathrm{lc}}$, the orbit is expected to proceed immediately to a smaller $r$, resulting in capture by the black hole. At the loss-cone boundary, the orbit of the test particle becomes marginally bound: the energy equals the effective potential $E^{2}-V(r)=0$. The effective potential exhibits an extremum at this radius, satisfying $\mathrm{d}V(r)/\mathrm{d}r=0$, with $\mathrm{d}^{2}V(r)/\mathrm{d}r^{2}<0$ indicating an unstable extremum.

By applying these conditions, the angular momentum of the loss-cone is related to the energy by~\cite{Sadeghian:2013laa}
\begin{align}\label{eq:unstable_L}
L_{\mathrm{lc}}^{2} = \frac{32 \bigl(G M_{\mathrm{BH}}\bigr)^{2}}{36E^{2} - 27E^{4} - 8 + E\bigl(9E^{2} - 8\bigr)^{3/2}}.
\end{align}
In the case where the test object comes from infinity, the energy of the test object is $E=1$ and $L_{\mathrm{lc}}=4GM_{\mathrm{BH}}$.

\subsection{Mapping from semi-major axis to energy-dependent $L_{\mathrm{lc}}$}\label{subsec:Energy_Dependent_Llc}
In the EMRI formation scenario, the test object cannot be treated as a freely falling system. As it traverses the relaxation-dominated region over many orbital cycles before entering the GW-dominated region, the test object experiences not only the gravitational field of the MBH but also perturbations from the surrounding field objects.

Once a test object enters the GW emission region, gravitational effects dominate. Then the relaxation effect can be neglected, and the test object can be treated as free falling. The initial energy is not unity but is determined by the curve $t_{\mathrm{gw}} = t_{\mathrm{rlx};\, r_{p}}$ in the $(1-e)$--$a$ plane.  

For an orbit, the radial velocity vanishes at both periapsis and apoapsis. Consequently, the right-hand side of Eq.~\ref{eq:EOM-schwarzshild} is equal to zero at these two points. Since $E$ and $L$ remain approximately constant over a single orbital period, evaluating Eq.~\ref{eq:EOM-schwarzshild} in periapsis ($r=r_p$) and apoapsis ($r=r_a$) yields two equations that can be solved for $E$ and $L$:
\begin{align}
    L^{2} = \dfrac{2 GM_{\mathrm{BH}} r_{a}^{2} r_{p}^{2}}{\,r_{a} r_{p} \left(r_{a}+r_{p}\right) - 2 GM_{\mathrm{BH}} \left(r_{a}^{2} + r_{a} r_{p} + r_{p}^{2}\right)\,},
\end{align}
\begin{align}
    E^{2} = \dfrac{(2GM_{\mathrm{BH}} - r_{a})(2GM_{\mathrm{BH}} - r_{p})(r_{a} + r_{p})}{r_{a} r_{p} (r_{a} + r_{p}) - 2GM_{\mathrm{BH}} \bigl(r_{a}^{2} + r_{a} r_{p} + r_{p}^{2}\bigr)}.
\end{align}
Here, $r_a = a(1+e)$ and $r_p = a(1-e)$ denote the orbital apoapsis and periapsis, respectively. Then, the energy can be simplified as

\begin{align}\label{eq:aetoE}
    E^2 = \frac{\left(a(1+e) - 2GM_{\mathrm{BH}}\right)\left(a(e-1) + 2GM_{\mathrm{BH}}\right)}{a\left[a\left(e^2 - 1\right) + GM_{\mathrm{BH}}\left(e^2 + 3\right)\right]}.
\end{align}

A test object with semi-major axis $a$ is driven by relaxation in the gravitational-wave emission regime with eccentricity $e(a)$. Consequently, imposing the condition $t_{\mathrm{gw}} = t_{\mathrm{rlx};\, r_{p}}$ yields the initial pair $(a, e(a))$. Using Eq.~\ref{eq:aetoE}, we determine the initial energy $E$ for which the test object can be treated as freely falling. Substituting this initial energy into Eq.~\ref{eq:unstable_L} then yields the loss-cone angular momentum at $a$. The loss-cone angular momentum $L_{\mathrm{lc}}$ is, in fact, a function of $a$.

\subsection{GR-corrected plunge pericenter}\label{subsec:GR_Corrected_Plunge}
The critical semi-major axis, $a_{\mathrm{cri}}$, substantially influences the event rate because it serves as the upper limit of the integral in Eq.~\ref{eq:evetrate}. As discussed previously, Eq.~\ref{eq:condition_to_form_emir} determines $a_{\mathrm{cri}}$. Consequently, $a_{\mathrm{cri}}$ is relevant to the plunge radius $r_{\mathrm{p};\,\mathrm{lc}}$.

For a Keplerian orbit, the squared orbital angular momentum is related to $a$ and $e$ by~\cite{Will:2012kq}
\begin{align}
L^{2} = GM_{\mathrm{BH}}a(1-e^{2}) = GM_{\mathrm{BH}}r_{p}(1+e).
\end{align}
Due to the extreme eccentricity of EMRIs, $e$ can be approximated as unity; therefore, $r_p = L^2/(2 G M_{\mathrm{BH}})$. Setting the angular momentum of the loss-cone at $4\,G M_{\mathrm{BH}}$ implies $r_{\mathrm{p};\,\mathrm{lc}} = 8\,G M_{\mathrm{BH}}$, as adopted in previous work. 

When the periapsis is sufficiently close to the MBH, the Newtonian approximation breaks down. Therefore, employing general relativity to compute the minimum pericenter distance is more appropriate. For a Schwarzschild MBH, by applying the unstable condition in Sec.~\ref{subsec:Schwarzschild_Llc}, the minimum pericenter distance for an orbit with angular momentum $L$ is given by~\cite{Sadeghian:2013laa}
\begin{align}
    r = \dfrac{6 G M_{\mathrm{BH}}}{1 + \sqrt{1 - 12 \left( \dfrac{G M_{\mathrm{BH}}}{L} \right)^{2}}}.
\end{align}

As discussed above, the test object cannot be treated as freely falling until its orbit enters the GW-emission region. The loss-cone angular momentum depends on the initial semi-major orbital axis. Consequently, $r_{\mathrm{p};\,\mathrm{lc}}$ is not constant but depends on $a$. Therefore, $a_{\mathrm{cri}}$ should be calculated by
\begin{equation}\label{eq:condition_to_form_emir2}
\begin{aligned}
t_{\mathrm{gw}} &= t_{\mathrm{rlx};, r_{\mathrm{p}}}, \\
a(1 - e) &=  \dfrac{6 G M_{\mathrm{BH}}}{1 + \sqrt{1 - 12 \left( \dfrac{G M_{\mathrm{BH}}}{L_{\mathrm{lc}}(a)} \right)^{2}}}.
\end{aligned}
\end{equation}
In practice, because EMRIs form within the influence radius of the MBH, $E$ does not deviate substantially from unity; therefore, $L_{\mathrm{lc}} = 4GM_{\mathrm{BH}}$ remains appropriate. Consequently, in a Schwarzschild background, one obtains $r_{p,\mathrm{lc}} = 4GM_{\mathrm{BH}}$.

\begin{figure}[t]
    \centering
    \includegraphics[width=1\linewidth]{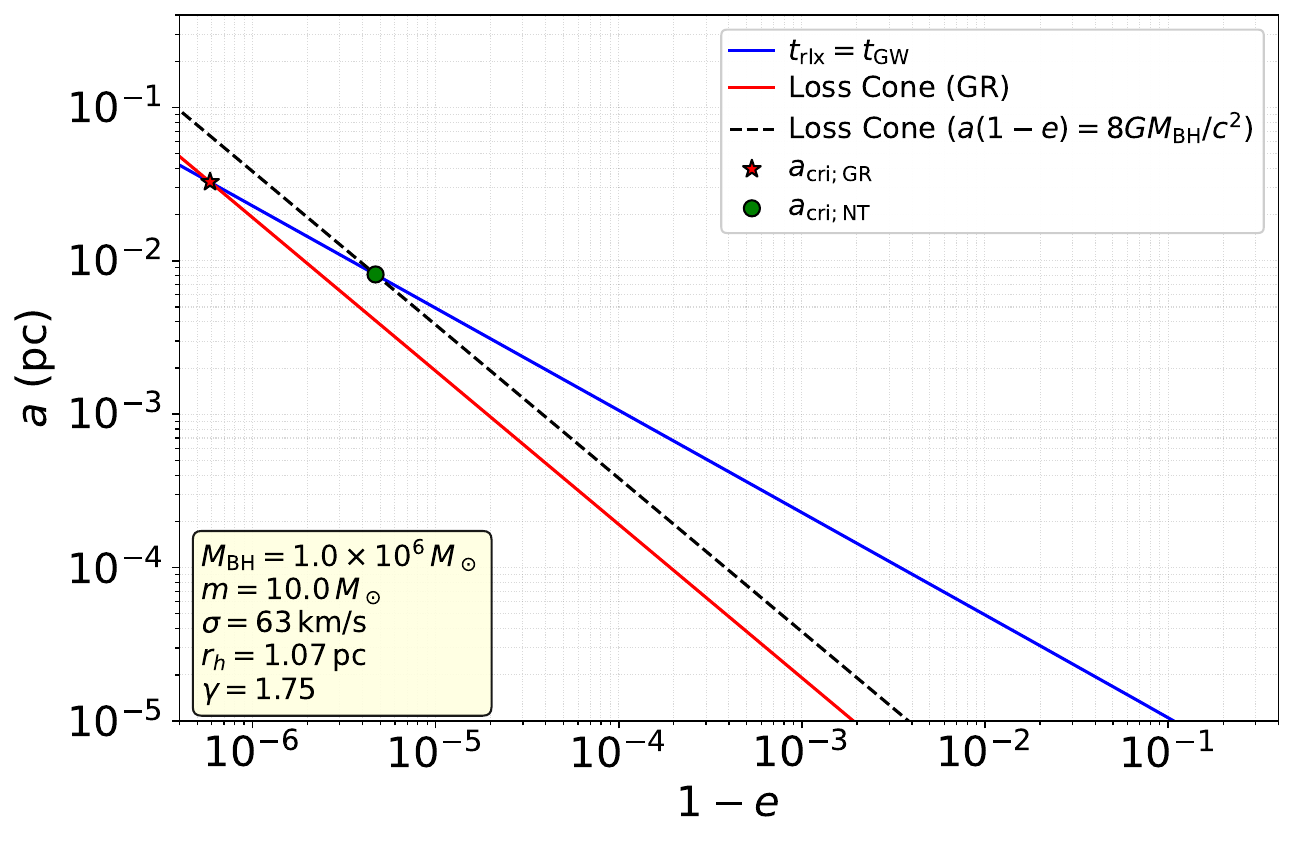}
    \caption{Comparison of two loss-cone curves in the $(1-e)$-$a$ plane. The black dashed line denotes the case $a(1-e)=8GM_{\mathrm{BH}}$, while the red solid line denotes the Schwarzschild-spacetime correction. The blue solid line indicates where the characteristic timescale of gravitational-wave emission equals the relaxation timescale for the evolution of periapsis. The circle- and star-shaped intersection points denote the values of $a_{\mathrm{cri}}$ corresponding to the two loss-cone curves, respectively.}
    \label{fig:acri}
\end{figure}

As shown in Fig.~\ref{fig:acri}, incorporating the Schwarzschild spacetime correction shifts the entire loss-cone boundary downward relative to the earlier result, which in turn increases $a_{\mathrm{cri}}$. With the parameters specified in Fig.~\ref{fig:acri}, before incorporating relativistic corrections, the critical semi-major axis was $8.15\times10^{-3}\,\mathrm{pc}$; after including these corrections, it increased to $3.26\times10^{-2}\,\mathrm{pc}$. $a_{\mathrm{cri}}$ is nearly increased by a factor of four, suggesting a pronounced effect on the event rate.

The extension to rotating black holes with Kerr spacetime introduces additional complexity. In such a case, the loss-cone boundary depends not only on energy but also on orbital inclination and spin of the black hole. Although closed-form expressions for the loss-cone are generally not available for eccentric inclined orbits in Kerr geometry, the relativistic framework developed here can be generalized using numerical methods~\cite{Amaro-Seoane:2012jcd}. Physically, compared to a non-rotating black hole, spin lowers the capture threshold for prograde orbits, effectively reducing the pericenter boundary and enlarging the phase space for EMRI formation. This shift in the loss-cone boundary would increase the critical semi-major axis and thereby enhance the overall EMRI event rate. 

\section{Effect on the EMRI event rate}\label{sec:Impact_on_EMRIs_event_rate}
In the vicinity of the MBH, we assume that the test objects and field objects have identical masses and that the test objects exhibit a quasi-steady, isotropic distribution in the semi-major-axis space, so the number within the interval from $a$ to $a+da$ is $N_{\mathrm{iso}}(a)\,da$. In the presence of two-body relaxation, the EMRI production rate per unit time over this interval can be expressed as~\cite{Amaro-Seoane:2019umn}
$$\frac{N_{\mathrm{iso}}(a)\,da}{\ln\!\bigl(L_{\mathrm{m}}/L_{\mathrm{lc}}(a)\bigr)\,t_{\mathrm{rlx}}(a)}.$$
If the cusp is partially depleted, it can regrow in the presence of MBH on a characteristic timescale $t_{\mathrm{cusp}} \approx 0.25\,t_{\mathrm{rlx}}$~\cite{Babak:2017tow}. During a duration $t_{\mathrm{cusp}}$, the number of test objects in the interval $[a,a+da]$ is replenished to $N_{\mathrm{iso}}(a)\,da$. Hence, the corresponding repopulation rate of test objects per unit time is
$$\frac{N_{\mathrm{iso}}(a)\,da}{0.25\,t_{\mathrm{rlx}}(a)}.$$
Therefore, the production of EMRIs at a given $a$ is limited by the smaller of the relaxation-driven capture rate and the cusp-repopulation rate. Integrating in the relevant semimajor axis range $[a_{\mathrm{min}},\,a_{\mathrm{cri}}]$, the total rate of EMRIs is 
$$\Gamma \;=\; \int_{a_{\mathrm{min}}}^{a_{\mathrm{cri}}} \min\!\left\{ \frac{N_{\mathrm{iso}}(a)\,da}{\ln\!\bigl(L_{\mathrm{m}} / L_{\mathrm{lc}}(a)\bigr)\,t_{\mathrm{rlx}}(a)} \;,\; \frac{N_{\mathrm{iso}}(a)\,da}{0.25\,t_{\mathrm{rlx}}(a)} \right\}.$$
This expression explicitly encodes the competition between inward diffusion into the loss-cone and the replenishment of the cusp population, ensuring that the EMRIs rate does not exceed the rate at which the reservoir of test objects is refilled.

The event rates $\Gamma$ in Table~\ref{tab:event_rate} show the Newtonian approximation with $L_{\mathrm{lc}}=4GM_{\mathrm{BH}}$ and $r_{\mathrm{p};\,\mathrm{lc}}=8GM_{\mathrm{BH}}$, and $\Gamma_{\mathrm{GR}}$ for the relativistic treatment in which $L_{\mathrm{lc}}$ is expressed as a function of the initial energy, and $r_{\mathrm{p};\,\mathrm{lc}}=6GM_{\mathrm{BH}}/\{1+[1-12(GM_{\mathrm{BH}}/L_{\mathrm{lc}})^2]^{1/2}\}$. As shown, the influence of including the Schwarzschild minimum radius becomes stronger as the profile slope $\gamma$ decreases. This occurs because a smaller slope implies a flatter stellar distribution, so for a fixed total number of objects, a larger fraction is located at greater radii. As a result, increasing $a_{\mathrm{cri}}$ leads to a more significant effect. Conversely, a large $\gamma$ produces the opposite trend. In addition, this influence appears to be insensitive to the MBH mass: for different $M_{\mathrm{BH}}$ values with the same $\gamma$, the event-rate ratio $\Gamma_{\mathrm{GR}}/\Gamma$ remains nearly unchanged.

The study in \cite{Babak:2017tow} has produced a catalog of EMRIs and estimated the number of EMRIs per year up to redshift $z=4.5$. The predicted events span more than three orders of magnitude, from $\sim$ 10 to 20,000. Since these numbers scale linearly with the assumed event rate, including relativistic corrections could raise the event rate by as much as an order of magnitude. As a result, depending on the specific model adopted, the expected number of EMRIs can range from about 10 up to $10^{5}$.

The relativistic corrections considered above address only the gravitational dynamics of the compact object itself. In reality, galactic nuclei may also host dark matter spikes surrounding the MBH~\cite{Gondolo:1999ef, Sadeghian:2013laa, Ferrer:2017xwm, Xie:2025udx}. The dark matter spike can alter the orbital parameters—including the pericenter—as the trajectory penetrates into the GW-emission region~\cite{Zhang:2024ugv, Yue:2024xhf, Chen:2025jch}. 
{We have therefore examined whether dark matter affects the event rate through its modification of the characteristic-timescale equality condition and the loss-cone boundary. For the former effect, we have already shown in our previous work that, for particle-like dark matter, the resulting impact on the event rate is negligible~\cite{Feng:2025fkc}. For the latter, The contribution is also subdominant for particle-like dark matter. The reason is that the loss-cone angular momentum and the plunge pericenter are primarily determined by the relativistic capture condition of the MBH. In the relevant inner region, the mass of dark matter enclosed within the orbit is much smaller than the MBH mass.}

 
\begin{table}[t]
\centering
\caption{EMRI event rates $\Gamma_{\mathrm{GR}}$ and $\Gamma$ for different MBH masses $M_{\mathrm{BH}}$ and stellar density slopes $\gamma$.}
\label{tab:event_rate}
\renewcommand{\arraystretch}{1.5}

\begin{tabular}{c@{\hspace{0.3cm}}ccccc}
\hline
$M_{\rm BH}(M_\odot)$ & $\gamma$& $\Gamma_{\mathrm{GR}}\,[\mathrm{yr}^{-1}]$ & $\Gamma\,[\mathrm{yr}^{-1}]$ & $\Gamma_{\mathrm{GR}}/\Gamma$ \\
\hline
\multirow{6}{*}{$10^4$} 
& $1/2$ & $9.58\times10^{-7}$ & $8.90\times10^{-8}$ &10.76 \\
& $1$ & $1.90\times10^{-6}$ & $2.31\times10^{-7}$ &8.23 \\
& $5/4$ & $2.98\times10^{-6}$ & $4.41\times10^{-7}$ & 6.75 \\
& $3/2$ & $5.27\times10^{-6}$ & $1.02\times10^{-6}$ & 5.18 \\
& $7/4$ & $1.15\times10^{-5}$ & $3.22\times10^{-6}$ & 3.57 \\
& $2$ & $4.17\times10^{-5}$ & $2.13\times10^{-5}$ & 1.96 \\
\hline
\multirow{6}{*}{$10^5$} 
& $1/2$ & $4.75\times10^{-7}$ & $4.42\times10^{-8}$ &10.76 \\
& $1$ & $9.83\times10^{-7}$ & $1.21\times10^{-7}$ & 8.15 \\
& $5/4$ & $1.60\times10^{-6}$ & $2.39\times10^{-7}$ & 6.70 \\
& $3/2$ & $2.96\times10^{-6}$ & $5.77\times10^{-7}$ & 5.14 \\
& $7/4$ & $6.91\times10^{-6}$ & $1.96\times10^{-6}$ & 3.53 \\
& $2$ & $2.85\times10^{-5}$ & $1.51\times10^{-5}$ & 1.89 \\
\hline
\multirow{6}{*}{$10^6$} 
& $1/2$ & $2.42\times10^{-7}$ & $2.27\times10^{-8}$ &10.65 \\
& $1$ & $5.20\times10^{-7}$ & $6.44\times10^{-8}$ & 8.08 \\
& $5/4$ & $8.73\times10^{-7}$ & $1.31\times10^{-7}$ & 6.66 \\
& $3/2$ & $1.68\times10^{-6}$ & $3.30\times10^{-7}$ & 5.08 \\
& $7/4$ & $4.13\times10^{-6}$ & $1.19\times10^{-6}$ & 3.48 \\
& $2$ & $1.92\times10^{-5}$ & $1.06\times10^{-5}$ & 1.81 \\
\hline
\end{tabular}
\end{table}

\section{Conclusion}\label{sec:conclusion}
We have presented a relativistically consistent analytic framework for estimating EMRI event rates in Schwarzschild spacetime by incorporating an energy-dependent loss-cone angular momentum into a one-dimensional Fokker–Planck treatment of angular momentum diffusion. Replacing the usual Newtonian, energy-independent loss-cone prescription and revising the plunge condition to use the innermost stable radius of the GR, enable us to calculate self-consistently both the critical semi-major axis and the effective loss-cone boundary. These improvements shift the loss-cone boundary inward and consequently increase the critical semi-major axis. As a result, the estimated EMRI rates are almost increased by one order of magnitude relative to Newtonian treatment. This enhancement becomes more pronounced for shallower stellar density slopes and depends only weakly on the MBH mass. Our results emphasize that relativistic capture physics is essential for reliable EMRI rate predictions and provide a straightforward path to include additional physical ingredients—such as MBH spin, orbital anisotropy, and secular torques—to further refine the forecasts for LISA and Taiji.

\begin{acknowledgements}
This work is partly supported by the National Key Research and Development Program of China (Grant No.~2021YFC2201901), 
and the Fundamental Research Funds for the Central Universities. 
\end{acknowledgements}

\bibliography{refs}{}

\end{document}